# Trade-off between reconstruction accuracy and physical validity in modeling turbomachinery particle image velocimetry data by physics-informed convolutional neural networks

Maryam Soltani [1], Ghasem Akbari [2,*], and Nader Montazerin [1]

[1] Department of Mechanical Engineering, Amirkabir University of Technology, Tehran, Iran
[2] Department of Mechanical Engineering, Qazvin Branch, Islamic Azad University, Qazvin, Iran
* Corresponding Author: g.akbari@qiau.ac.ir

## Abstract

Particle image velocimetry (PIV) data is a valuable asset in fluid mechanics. It is capable of visualizing flow structures even in complex physics scenarios, such as the flow at the exit of the rotor of a centrifugal fan. Machine learning is also a successful companion to PIV in order to increase data resolution or impute experimental gaps. While classical algorithms focus solely on replicating data using statistical metrics, the application of physics informed neural networks (PINN) contributes to both data reconstruction and adherence to governing equations. The present study utilizes a convolutional physics-informed auto-encoder to reproduce planar PIV fields in the gappy regions while also satisfying the mass conservation equation. It proposes a novel approach, which compromises experimental data reconstruction for compliance with physical restrictions. Simultaneously, it is aimed to ensure that the reconstruction error does not considerably deviate from the uncertainty band of the test data. Turbulence scale approximation is employed to set the relative weighting of the physical and data-driven terms in the loss function to ensure that both objectives are achieved. All steps are initially evaluated on a set of direct numerical simulation data to demonstrate general capability of the network. Finally, examination of the PIV data indicates that the proposed PINN auto-encoder can enhance reconstruction accuracy by about 28% and 29% in terms of mass conservation residual and velocity statistics, respectively, in expense of up to 5% increase in the number of vectors with reconstruction error higher than the uncertainty band of the PIV test data.

# I) Introduction

The world of fluid mechanics is faced with huge numerical and experimental data sets which makes it a perfect environment for the application of artificial intelligence tools. Extracting models from data in this field requires an expert who can manage the process and make crucial decisions [1]. The advancement of deep learning techniques has led to active research in applying them to the development of innovative approaches in the field of fluid mechanics [2,3].

Direct numerical simulation (DNS) or high-order developed codes in computational fluid dynamics (CFD) are accurate methods which are continuously developed for determining the state of the flow field. Results have shown that recent methods are not always cost-effective in terms of time and computational resources and require significant time and expense to obtain their solutions in large fields.

Implementation of experimental methods helps to identify complicated flow fields in which numerical approaches may fail due to enormous computational cost. Although experimental data is aligned with the physical reality of the phenomena, they may be contaminated with some noise and measurement uncertainty. Among various experimental methods, particle image velocimetry (PIV) and its derivatives play a special role in spatio-temporal analysis of turbulent fields. Presence of some clustered or spot-wise missing data, induced by optical noise or inadequate seeding density, are challenges that should be resolved prior to further analysis. Different methods are employed for gappy velocity field reconstruction [4,5]. Classical methods for this reconstruction are statistical, but with the advancement of machine learning algorithms and deep learning, the idea of utilizing tools such as KNN (K-nearest-neighbor) or neural networks [6-8] for data imputation are proposed. A wide variety of deep learning methods, such as convolutional neural networks [9] and generative adversarial networks [10,11], are applied for the reconstruction of two-dimensional images and enhance the resolution of flow fields [12-14]. The common issue with these methods is that although they eventually report a numerical value for the velocity vector, the result only attempts to reproduce the experimental data without considering the physics of the problem.

Reconstructed fields from conventional methods for data completion, deviate from expected governing equations [15]. The governing physical equations of each problem could act as constraints to produce a more physical reconstruction. In 2019, the idea of physics-informed neural networks (PINN) was proposed [16]. PINN fundamentally leverages traditional neural networks and incorporates problem-dependent loss functions to address typical loss functions and also minimize deviations from the governing equations. This quickly sparked interest in applying these networks to different applications such as fluid mechanics for determining the velocity and pressure fields in conjunction with the Navier-Stokes equations [17]. This idea incorporated information derived from physical laws and reduced the need for big data which is not always available [18].

PINN is numerously tested on DNS results to show its capabilities in applications where the data is ill-posed or incomplete [19-21]. The main power of PINNs lies in having robust loss functions that incorporate the physics of the problem. Various choices of such loss functions are made in different research studies [21-25]. In each of these studies, the loss function comprises of two components. The first component, termed the data-based term, aims to minimize the error between the predicted quantities and the available data. The second component, known as the physics-informed part, is formulated based on the inherent nature of the problem. Adding a regularization term to the loss function of a neural network involves introducing extra weights or hyper-parameters that need to be adjusted. While the significance of these parameters was previously unknown, and they were often chosen without explicit rationale in earlier works, recent studies have introduced diverse methodologies. These include analyzing the Hessian of the loss function [26], utilizing neural tangent kernel theory [27], and integrating the weighting hyper-parameters alongside trainable parameters [28].

The application of a physics-informed neural network to experimental fluid mechanics data, which might be sparse and subject to limitations imposed by the data acquisition procedure, is promising [29]. For instance, PINN is applied to experimental datasets from tomographic background oriented schlieren to infer velocity and pressure fields from temperature data [30]. This aids in reconstruction of dense velocity and pressure fields from sparse experimental data [31], as well as the reconstruction of velocity in one direction with the known velocity in another direction within a two-dimensional domain [22]. Integration of both the continuity and momentum equations into the physical loss term requires the availability of time-resolved three-component velocity field within a volume, enabling the calculation of all instantaneous velocity gradients. Although such measurement data and PINN reconstruction is possible using modern measurement techniques like time-resolved tomographic particle tracking velocimetry [25,32] it is not affordable for many research teams round the globe. This challenge presents a significant opportunity for research, where less complete data, such as single-plane PIV data, can be enhanced to higher levels of completeness and accuracy using physics-informed machine learning.

It is essential to note that in comparison with the reconstruction of DNS data, in the application of PINN to experimental data, there are more significant errors in the collected data points. The contribution of this paper lies in its consideration of a trade-off, within the confines of experimental error, wherein reconstructed data may deviate from the original field but exhibits improved adherence to the governing equations. The current study utilizes single-plane PIV data as the input data and a residual indicator of the out-of-plane velocity gradient (derived from the general three-dimensional continuity equation) as the physical loss term. Such reconstruction completes the missing in-plane vector maps and also achieves better consistency with the mass conservation equation. This approach initially aims to perform reconstruction on artificially ill-posed DNS data and then applies the algorithm to the challenging problem of PIV data of a high-Reynolds turbulent flow at the rotor exit of a centrifugal fan. The goal is to show the trade–off between reconstructing the measured velocity field and complying with the mass conservation equation. The article further contributes by examining the impact of using weighted coefficients for different terms of the loss function on satisfying the mass conservation equation.

The rest of this article is organized as follows. In the next section, PIV data and the required preprocessing steps are introduced. It follows by the proposed methodology for the PINN and configuration of the network and loss function. The fourth section discusses the results for reconstruction of DNS and PIV data, providing a detailed evaluation of the influence of weighting coefficients on the accuracy of the proposed PINN network.

## II) Experimental data and pre-processing

### 2-1- PIV dataset

Stereoscopic PIV is used to measure velocity components at the rotor exit region of a forward-curved centrifugal fan. Data is acquired at an encoded position of the rotor with the width of $165 mm$, and inner and outer diameters of $285 mm$ and $350 mm$, respectively. The steady rotational speed of the rotor is 745 rpm and the machine Reynolds number is $3.18 \times 10^5$. Detailed specifications about the fan geometry and operating conditions are presented in [33]. The experimental setup is shown in Figure 1 and includes the following components [33]:

- A 43-blade centrifugal fan constructed from galvanized plate
- A test setup and an outlet duct compatible with the ISO 5801 standard [34]
- A volute made of transparent Plexiglas to provide optical access to the test section
- An optical encoder to facilitate phase-locked encoding of rotor orientation
- A double-cavity Quantel Brilliant Nd-YAG laser equipped with an optical guide system that delivers the laser sheet to the test section
- Two FlowSense® 1600×1186-pixel double-frame CCD cameras
- SAFEX F2010 plus fog generator
- Dantec FlowMap® system for synchronization of laser, camera, and encoder actions.

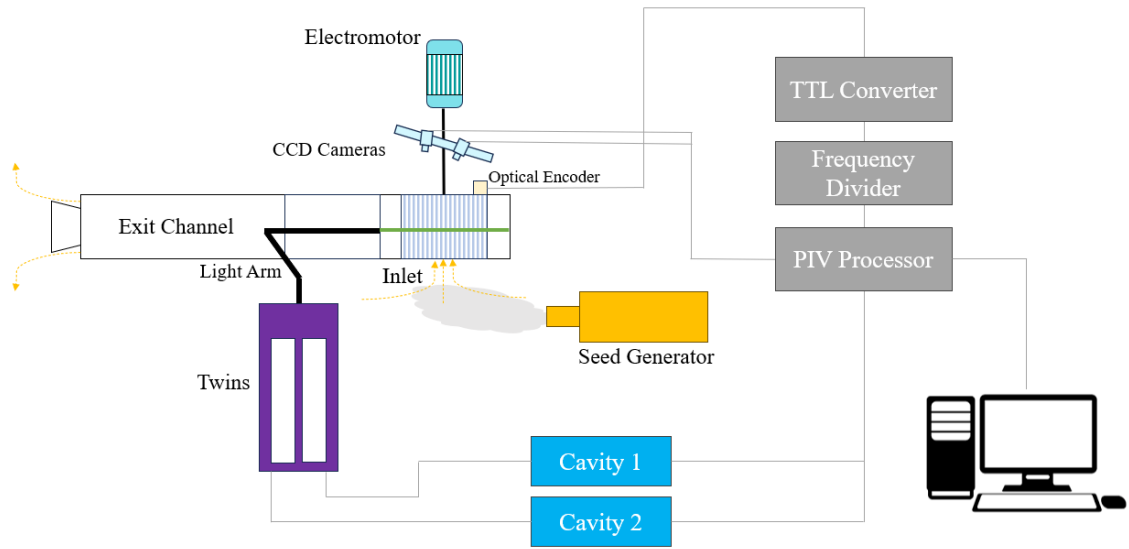

Figure 1 Schematic representation of the SPIV System

For acceptable tracing accuracy, the particle response time should be considerably smaller than the smallest time scale of the flow, i.e. the Kolmogorov time scale. In other words, the Stokes number must be significantly smaller than unity. Based on scale analysis, the Stokes number of the measurements was estimated to be $2 \times 10^{-5}$ [33], that confirms that the implemented seeding particles are appropriate for fluid flow tracking. The time interval between consecutive laser pulses for capturing two successive frames of each camera is 20 µs.

Regrading the resolvable flow structures and PIV resolution, it is known that setting the value of the smallest resolvable velocity structure in the range of $20\eta - 90\eta$ covers 95-65 per cent of turbulent kinetic energy ($\eta$ is the Kolmogorov length scale, which was estimated to be 57 µm)[33]. The smallest resolvable velocity structure is set about $35\eta \approx 2mm$. Considering the camera magnification factor, pixel pitch of the camera CCD and based on 64×64 pixel interrogation area, the dimension of FOV is calculated to be $5.36 \times 3.97 cm^2$.

Velocity measurements are conducted in the upper rotor region at a plane perpendicular to the rotor axis, specifically at Z/B = 0.4 [33]. Here, Z represents the axial coordinate, measured from the volute inlet, while B is the width of the volute (Figure 2(b)). Ten adjacent fields of view (FOVs) with shared boundaries are selected to cover a larger measurement area, as illustrated in Figures 2(a) and 2(c). One with slight overlap with the rotor region (FOV 1), and a 3 by 3 grid of FOVs outside the rotor region (FOVs 2-10). Our concern in this study is to evaluate how the new PINN network acts in reconstruction of velocity field for the rotor exit region. Part (d) of Figure 2 illustrates phase-average of the velocity field over the ten overlapping areas, and indicates that flow is characterized by severe interactions among jet and wake flow structures. Such interaction is more pronounced in the near-rotor regions, i.e. FOVs 1-4. Among the 10 measured areas, FOV 1 is partly obstructed by the rotor, and FOVs 2-4 exhibited almost similar behavior in data reconstruction. Therefore, FOV 2 is selected as the target region for the present study, that is a suitable representative with considerable jet/wake interactions and without optical obstruction by the rotor.

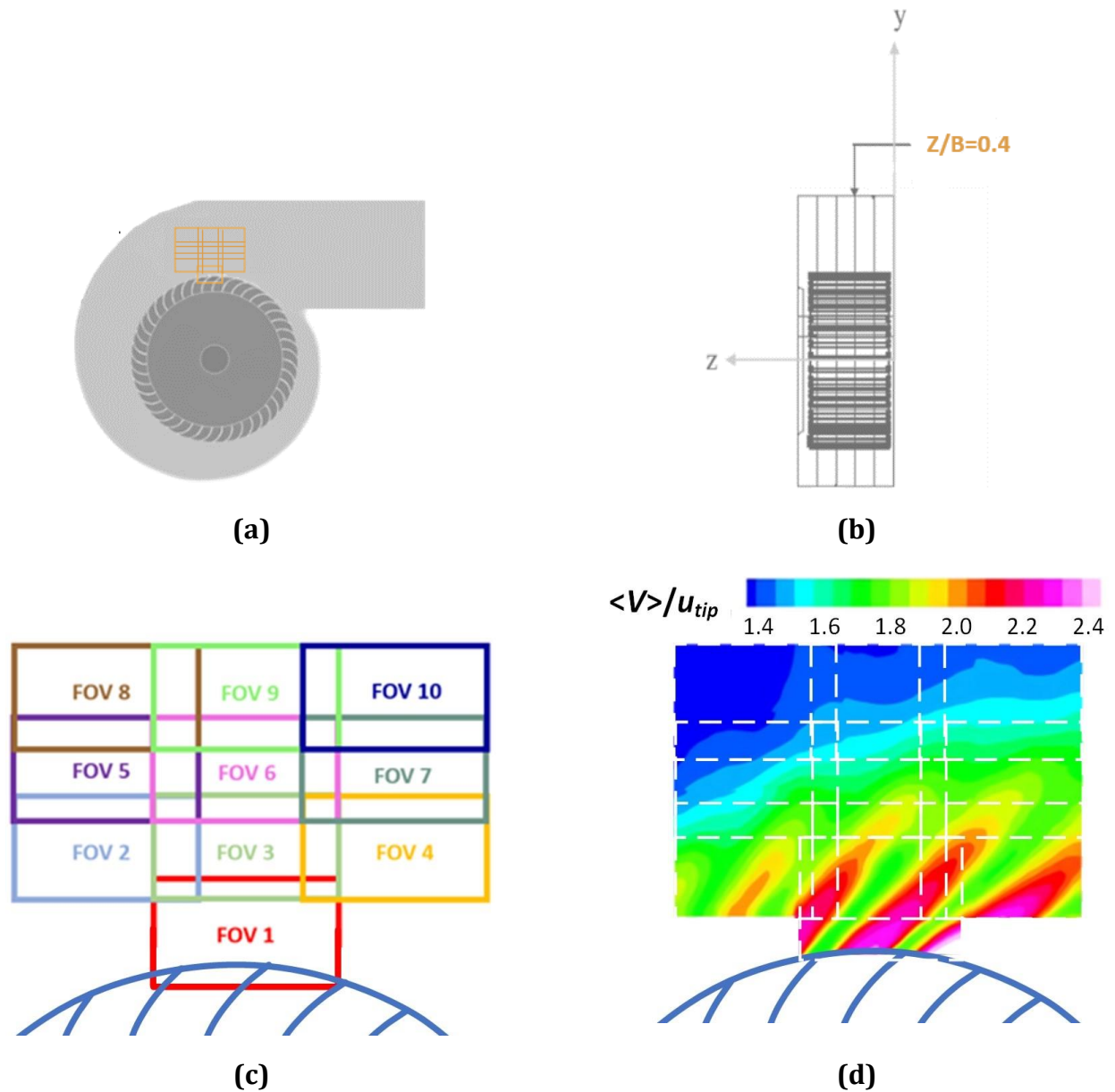

Figure 2 Sketch of centrifugal fan, and position of FOVs along with their velocity field at the rotor exit region: (a) front view of the rotor; (b) side view of the rotor; (c) position of 10 original FOVs; (d) phase-averaged velocity over the 10 FOVs.

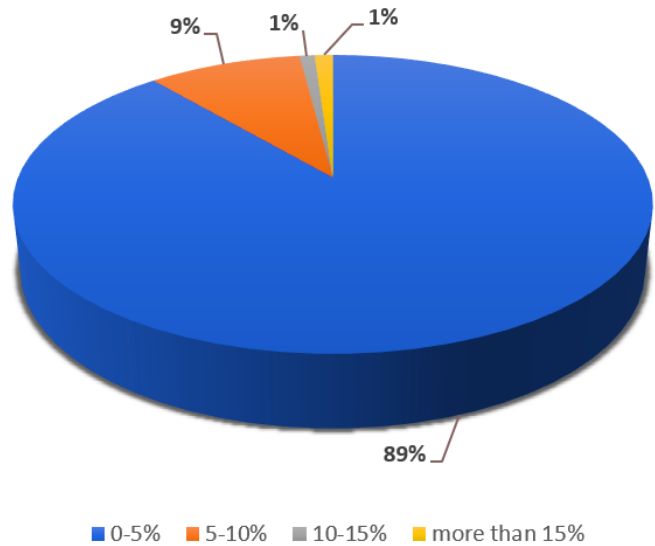

Figure 3 Percentage of PIV data gaps in FOV 2.

The number of snapshots for each FOV is 2500. During the data validation steps, some data is removed leading to generation of some gappy points in each snapshot. These invalidated data is due to lack of sufficient and homogeneous seeding, optical noises, loss-of-pair effect and complexity of the flow field. There is no general pattern for the location and the size of the gaps. In the present study, these gaps are classified into two groups. The first one are clustered gaps with more than one missing data in the neighborhood of each point, and the second one are individual gaps where the gappy point is surrounded from top, down, left and right neighbors by available data. In addition to these clustered or individual gaps (termed also as the original gaps), some artificial gaps are generated by a random algorithm. The purpose of these artificial gaps is to provide some test data for comparison of the model's reconstruction with the original velocity at those points.

Figure 3 illustrates the percentage of gaps in snapshots of FOV 2, and it indicates that 98% of the images have at least 90% valid data. Among all FOVs, the highest gap percentage belongs to FOV 1, due to some optical blockage with the rotor region. FOVs 5-9 which are farther away from the rotor, produce superior data quality due to more homogeneous seeding distribution and moderate velocity gradient.

## 2-2- Pre-processing

Preprocessing is applied to raw PIV data prior to velocity field reconstruction using PINN network. The resulting velocity field in some snapshots of the PIV experiments lacks a discernible pattern due to insufficient seeding particles. In other words, the marginal difference between the values of velocity vectors, results in a field with no discernible gradient that could avoid identification of diverse flow patterns. Snapshots with the standard deviation of all normalized velocity vectors smaller than 1.5 are therefore identified and excluded from the dataset, before training the model.

Next, outliers in each snapshot are detected and replaced with gaps based on the z-score. This factor relies on the mean ($\mu$) and standard deviation ($\sigma$) of velocity field in each snapshot. If any of the velocity vectors is not within the range $[\mu - 3\sigma, \mu + 3\sigma]$, it is identified as an outlier and removed.

The Next pre-processing step is normalization of the snapshot data. The global maximum and minimum velocities among all snapshots ($V_{min}$ and $V_{max}$) are used for normalization, as follows:

$$V^* = \frac{V_{original} - V_{min}}{V_{max} - V_{min}}. \tag{1}$$

Random artificial gaps are then added to each snapshot, to provide regions with true data for testing the model. The next step is initial imputation of the original and artificial gaps with a classic statistical method, in order to provide a complete vector map as the input for the PINN network. This is because the convolution kernel embedded in the PINN model cannot be applied to a sub-region containing missing values. Two-dimensional cubic spline is utilized to impute the data at the gap regions for each snapshot [15]. The PINN architecture then modifies this initialization, and makes the reconstruction more mature by implementing both data-driven and physics-informed metrics, as discussed in Section III.

## III) Methodology

PINNs are neural networks where both data and equations that are expected to govern that data are present in the loss function. Figure 4 provides a general overview of physics-informed neural networks where the selection and form of input can vary depending on the problem and the type of available data.

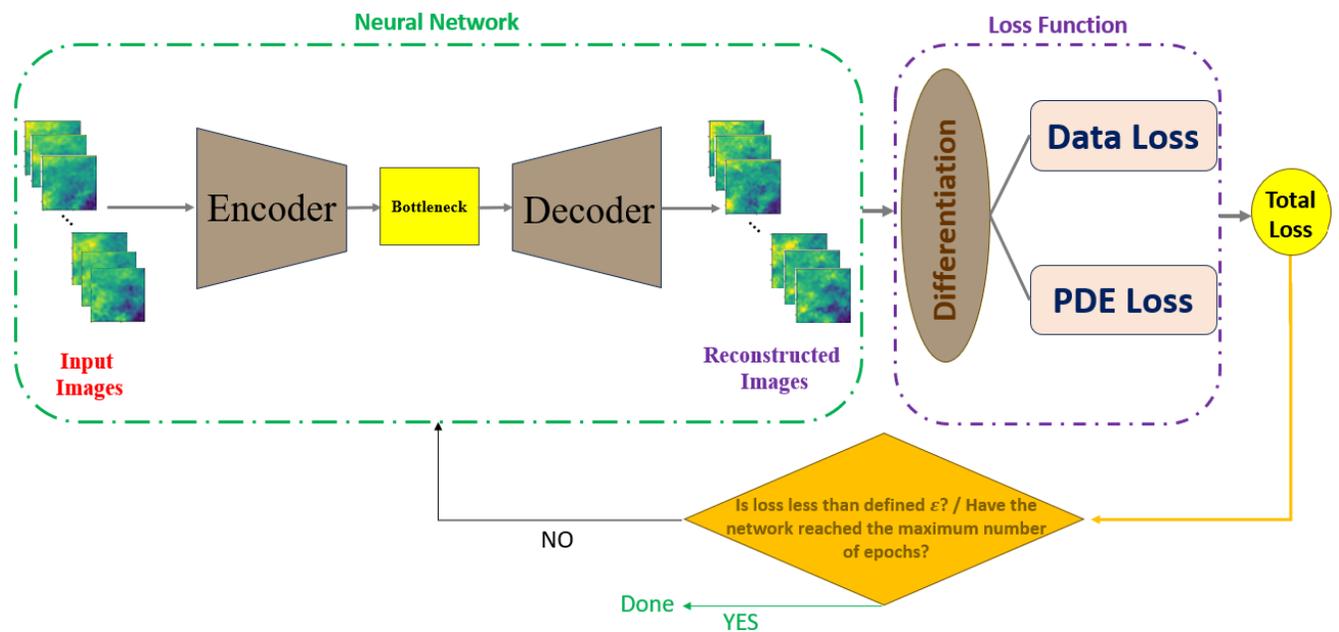

Figure 4: General flow of the physics-informed convolutional auto-encoder

## 3-1- Network architecture

Previous research on PINNs frequently selected spatial coordinates and time stamps as inputs for the network [30]. Such input is possible in the case of DNS or synthetic data, with sufficient spatio-temporal resolution. However, the PIV data is sparse, especially in the time domain, and therefore utilization of spatial coordinates and time as inputs does not allow automatic differentiation. In such circumstance, using velocity snapshots at different time stamps as individual input images and applying a convolutional network to identify spatial dependencies is the reasonable choice.

A convolutional encoder-decoder network is designed for this purpose, that receives a two-channel image (two feature maps corresponding to two in-plane velocity components) as the input and after imputation of the gaps, reconstructs the output vector map with the same resolution. Figure 4 schematically represents the encoder-decoder implementation in the current study. The details of the network is presented in Figure 5 for two-channel images with dimension 36×36 vector maps for PIV data. The size of convolution kernels is $3 \times 3$, and the same padding is used in convolution kernels to avoid any problem in the boundaries of images. Rectified linear unit (ReLU) as a widely-used and computationally-efficient activation function is used after each layer, to mitigate the vanishing gradient problem in the training procedure [35].

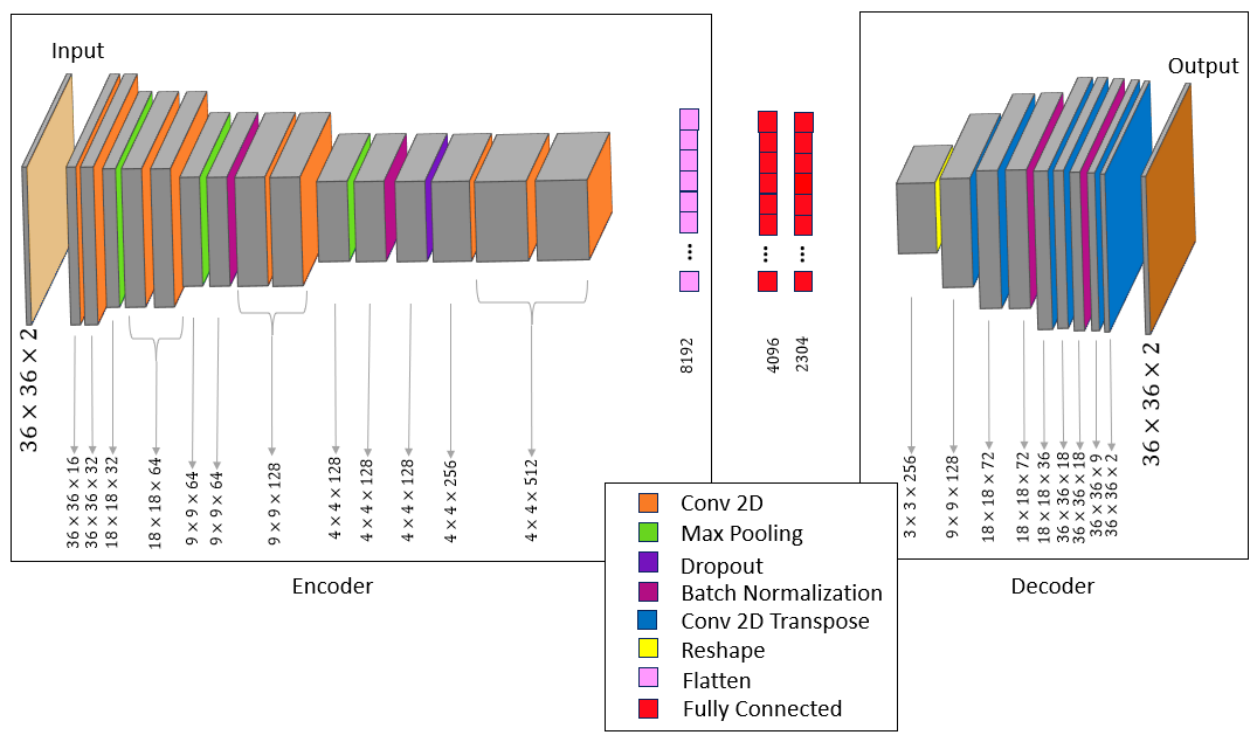

Figure 5 Architecture of the proposed physics-informed auto-encoder for PIV Data

The chosen network optimizer for the PINN model can significantly impact its performance and speed [36]. ADAM optimizer is used in the current research. At initial epochs, learning rate is constant, but an exponential decaying function is used to decrease learning rate in each epoch from initial value of 0.001. The network is designed based on the open-source platform TensorFlow v2.15.0, and computations are preformed through GPU of Google Colaboratory Pro with an NVIDIA Tesla P100 or T4 GPU (16 GB VRAM), Intel Xeon CPU (2 cores, 2.3 GHz), 25 GB RAM, and 150 GB temporary disk storage.

## 3-2-Loss Function

A physics-informed neural network constrains the network to learn specific equations that encapsulate the physics inherent to the problem. The idea of adding physics-informed loss function to convolutional neural network for enhanced reconstruction is introduced to match the patterns with the physics of the flow [37,38]. PINN combines conservation equations for a more realistic prediction. In fact, different equations (even with different dimensions) can be involved to report a final metric to address both velocity data and fluid dynamics equations [22,25]. The loss function in such networks includes terms that are dependent on the applied physics and other similar factors, thereby guiding predictions beyond mere replication of the original values.

The classic data-based loss function evaluates the error between model prediction and actual values:

$$L_{data} = \frac{1}{N_{data}}\left(\sum_{j=1}^{N_{data}} |U_{model} - U_{actual}|^2\right) \tag{2}$$

where $N_{data}$ is the number of evaluation samples and $U$ represents the velocity vector. For the physics-based loss function, various equations can be employed depending on the nature of the problem. Generally, it can be formulated with a residual that is calculated based on the physical equation ($R$), as follows:

$$L_{physical} = \frac{1}{N_{ph}}\sum_{j=1}^{N_{ph}}|R_{model} - R_{actual}|^2 \tag{3}$$

where $R_{model}$ and $R_{actual}$ are the residuals evaluated based on the model prediction and actual data, respectively. $N_{ph}$ represents the number of samples participating in calculation of the residual. There is no requirement for the same value of $N_{ph}$ and $N_{data}$. The following weighted sum can be applied to adjust the impact of data-based and physics-based loss terms:

$$L_{tot} = \alpha_{data}L_{data} + \alpha_{physical}L_{physical} \tag{4}$$

where the parameters $\alpha_{data}$ and $\alpha_{physical}$ are weight coefficients.

When incorporating physical constraints, particularly those derived from experimental datasets, the complexity of the optimization process increases significantly. The neural network must now be optimized based on multiple objectives of different nature and value: a combination of statistical data-driven metrics and physics-based criteria. This dual optimization is challenging because experimental datasets often contain systematic and random measurement errors, leading to deviations from the idealized physical laws. In such a scenario, it is not always feasible to satisfy both statistical metrics and physical criteria concurrently and strictly. Excessive focus on optimizing the network architecture for maximum reconstruction accuracy can lead to overfitting, thereby compromising the model's ability to generalize to new unseen data.

To address this challenge, we designed a cost function that strikes a balance between effective reconstruction of the flow field and adherence to physical laws. In other words, whether it is possible to improve both of these aspects in the empirical data or if there is a compromise between them, much like a trade-off, where data-based term can vary within the experimental error band in order to bring the predicted data closer to the governing equations. The proposed methodology for calculating the physics-based loss and the weight coefficients is discussed in the following.

### 3-2-1- Physics-based loss function

Mass conservation law is considered as the physical constraint for the PINN model in the present study. Generally, the conservation of mass equation is expressed as follows:

$$\frac{\partial u}{\partial x} + \frac{\partial v}{\partial y} + \frac{\partial w}{\partial z} = 0 \tag{5}$$

where $u$ and $v$ are the in-plane velocity components and $w$ is the out-of-plane component. If the flow is two-dimensional (like the DNS dataset utilized in section 4-1), $\frac{\partial u}{\partial x} + \frac{\partial v}{\partial y} = 0$ reflects the mass conservation law perfectly. However, for a three-dimensional data (like the experimental dataset of the present study), $\frac{\partial w}{\partial z} \neq 0$. For the single-plane PIV data, $\frac{\partial w}{\partial z}$ can not be directly evaluated, and just the in-plane components of velocity gradient are measured. However, the out-of-plane component of velocity gradient can be reconstructed from the 3D continuity equation, based on the available in-plane components, i.e. $\frac{\partial w}{\partial z} = -\left(\frac{\partial u}{\partial x} + \frac{\partial v}{\partial y}\right)$. In this way, the term $R = \frac{\partial u}{\partial x} + \frac{\partial v}{\partial y}$ is considered as the representative (non-zero residual indicator) of the mass conservation law in the physical loss term. The value of the third velocity gradient for both the model and target (actual value) is calculated based on the definition of this redidual indicator. Specifically, the modeled in-plane velocities produced by the neural network are used to calculate $R_{model}$, while the measured in-plane velocity components are employed to calculate $R_{actual}$. The difference $R_{model} - R_{actual}$ is then incorporated into the definition of the physical loss term, as presented in equation (6). In this way, it evaluates how close the network can repeat the value of the third velocity gradient on the basis of the 3D continuity equation.

$$L_{physical} = \frac{1}{N_{ph}} \sum_{j=1}^{N_{ph}} \left| \left(\frac{\partial u}{\partial x} + \frac{\partial v}{\partial y}\right)_{model} - \left(\frac{\partial u}{\partial x} + \frac{\partial v}{\partial y}\right)_{actual} \right|^2 \tag{6}$$

The derivatives in equation (6) are calculated by second-order central difference scheme[39].

### 3-2-2- Data-based loss function

The initial idea for defining the data-based loss function consisted of five terms:

$$L_{data} = \frac{1}{N_d}\left(\sum_{j=1}^{N_d}|(u)_{model} - (u)_{actual}|^2 + \sum_{j=1}^{N_d}|(v)_{model} - (v)_{actual}|^2 \right.$$
$$+ \sum_{j=1}^{N_d}\left|\left(\frac{\partial u}{\partial x}\right)_{model} - \left(\frac{\partial u}{\partial x}\right)_{actual}\right|^2$$
$$+ \sum_{j=1}^{N_d}\left|\left(\frac{\partial v}{\partial y}\right)_{model} - \left(\frac{\partial v}{\partial y}\right)_{actual}\right|^2 + \sum_{j=1}^{N_d}\left|\left(\frac{\partial v}{\partial x} - \frac{\partial u}{\partial y}\right)_{model}\right.$$
$$\left.\left. - \left(\frac{\partial v}{\partial x} - \frac{\partial u}{\partial y}\right)_{actual}\right|^2\right) \quad (7)$$

The first two terms in the right-hand-side of Equation (7) pertain to the velocity prediction error in horizontal and vertical directions. The third and fourth terms evaluate the predicted in-plane velocity gradients with the actual values, and the last term assesses reconstruction of the out-of-plane vorticity component (the only calculable component for a single-plane PIV data). The inclusion of the last three derivative-based terms significantly amplifies numerical errors in the loss function. This in turn, slows down convergence, necessitates utilization of more epochs, and increases the risk of overfitting. So, the final form of data-based loss function applied in the present study is as follows:

$$L_{data} = \frac{1}{N_d}\left(\sum_{j=1}^{N_d}|(u)_{model} - (u)_{actual}|^2 + \sum_{j=1}^{N_d}|(v)_{model} - (v)_{actual}|^2\right) \quad (8)$$

Finally, contribution of the physics-informed loss (equation (6)) and the simplified data-driven loss (equation (8)) can be adjusted using the weighting factors, based on equation (4), to calculate the total loss.

Another important aspect is the determination of number of locations that participate in each term of the loss function, i.e. $N_{ph}$ and $N_{data}$. Calculation of the derivatives in the physical term at each point requires velocity data at the four neighbors of that point. Evaluation of the generated artificial gaps indicated that the number of points satisfying such condition was not sufficient to obtain a stable loss. So, for the physical term, every point in the field with all four neighbors available, is employed in the calculation of loss function. On the other hand, the data-based loss term does not require knowledge of the neighbor data, and it is only calculated at the artificial gaps.

### 3-2-3- Weight coefficients

The parameters $\alpha_{data}$ and $\alpha_{physical}$ in Equation (4) determine the influence of different terms in the loss function, which has not been explicitly discussed in initial research on physics-informed neural networks [16]. Although in some studies, changes in these weights significantly changed the outcomes, but no specific rationale for selecting final values was put forward [29-31]. The importance of these weighting coefficients is crucial for accelerating network convergence and also for providing a logical definition of the relationship between different terms. Ideas for how to select or update these coefficients are suggested in some papers, but these ideas lack a clear physical connection to the type of flow [27,40]. Although these methods lack a physics-based justification, they show significant improvements compared to cases where the loss function is considered with equal weighting coefficients.

It is possible to dynamically update these coefficients during the training process. Adaptive weighting using loss ratio and meta-learning are two methods for such purpose. The idea in the former method is to adjust the coefficents $\alpha_{data}$ and $\alpha_{physical}$ in a way that keeps the contributions of data-driven and physics-based terms balanced in each epoch of the training process. In the meta-learning approach, $\alpha_{data}$ and $\alpha_{physical}$ are treated as learnable parameters, updated implicitly along with the network weights during each epoch. While these dynamic adjustment strategies can indeed balance the contributions of the loss function components, they also introduce potential numerical instabilities. This is particularly relevant in our study, where the data-driven and physics-based terms are derived from experimental data that inherently contains some degree of uncertainty. Such instabilities could adversely affect the convergence and overall performance of the model. To mitigate these risks, we have opted for a more stable approach based on turbulence scale analysis, proposed by Leoni et al. [25], to approximate the order of magnitude of weight coefficients. In addition to more computational stability, the advantage of this method is that it pays attention to physical nature and turbulence characteristics of the flow.

The following estimation is suggested for the ratio of weight coefficients (γ), which is based on the definition of total loss (Equation (4)) assuming the same order of magnitude for data-based and physics-based terms:

$$\gamma = \frac{\alpha_{data}}{\alpha_{physical}} \sim \frac{L_{physical}}{L_{data}} \tag{9}$$

Based on Equations (6) and (8), $L_{physical} \sim (\frac{\partial u}{\partial x})^2$ and $L_{data} \sim u^2$. Since the velocity field is normalized, the order of $u$, and therefore $L_{data}$ is unity, leading to $\gamma \sim (\frac{\partial u}{\partial x})^2$ from Equation (9). The order of magnitude for the velocity gradient is estimated for a turbulent flow with turbulence dissipation $\varepsilon$ and length scale $l$ as follows [25]:

$$\frac{\partial u}{\partial x} = \varepsilon^{1/3} l^{-2/3}. \tag{10}$$

The largest value of this estimate happens at the smallest scale of turbulent flow, i.e. the Kolmogorov length scale $\eta \sim \nu^{3/4} \varepsilon^{-1/4}$. So, equation (10) is expressed as follows:

$$\frac{\partial u}{\partial x} = \varepsilon^{1/2} \nu^{-1/2} \tag{11}$$

According to PIV data of the presents study and using equation (11), $\frac{\partial u}{\partial x}$ is estimated to be about 10, and therefore the ratio of weight coefficients is estimated to be about 100.

### 3-3- Network Assessment

The loss function should align with a suitable metric for evaluating network performance. In the present study, evaluation is done by plotting predicted values for the gaps vs. the original values expected in a line graph, with the same scales for the horizontal and vertical axes. Accordingly, a 45-degree line will be the reference line for network evaluation and there is no error in the reconstruction for data perfectly aligned with this reference line. A set of unseen snapshots is used for assessment. The mean absolute error ($MAE$) between prediction and target values (average vertical distance between cluster of points and the reference 45-degree line) is considered as the measure of error in order to enable comparisons across different snapshots:

$$MAE(\phi) = \frac{1}{N} \sum_{i=1}^{N} |\phi_{predicted} - \phi_{target}| \tag{12}$$

where $\phi$ represents each of the target quantities for evaluation, including horizontal and vertical velocity components ($u$ and $v$) and mass conservation residual ($MC$), defined as follows:

$$MC = \frac{\partial u}{\partial x} + \frac{\partial v}{\partial y} \tag{13}$$

## IV) Results and Discussion

The idea of establishing a trade-off between reproducing the exact velocity pattern of a flow, or missing the original data within its experimental error band but complying with the mass conservation law is examined in the present article with two data sets. Initially, a DNS dataset is used to allow evaluation of the network on a data with no gaps and negligible uncertainties in comparison with experimental data. It is expected that addition of the mass conservation equation (physical term) to the loss function, would reduce errors and help velocity field reconstruction. Then a PIV dataset with an estimated uncertainty level is used to evaluate the model performance. The addition of the physical term in this case could improve data reproduction or bring about a trade-off between data deviation from its original values for improved adherence to the governing equation.

## 4-1-DNS data reconstruction

DNS data is the result of direct simulation of the Naiver-Stokes equations (including mass conservation law) and follows them strictly. There is no error band (except for the very tiny computational errors) to allow for a trade–off towards improved adherence to the mass conservation law. It is therefore expected that the addition of the mass conservation equation to the loss function should bring the reconstructed pattern closer to the original data and reduce the reconstruction error.

A set of 150 time-resolved snapshots from a 2D DNS data around a cylinder [41,42] is used as the test case. In comparison with the PIV data, DNS data needs a minimal pre-processing due to negligible uncertainties, the absence of outliers, high quality of velocity data and availability of all data points. About 75% of the data is used for training and the remaining for the test procedure.

Velocity values in all snapshots are initially normalized and random vector maps in each image are transformed into individual or cluster gaps (about 25% of data at each snapshot is transformed to gaps). Gap positions are different for various snapshots, but they are kept the same at different executions. The next step is the imputation of each removed data point using cubic spline method. This is done just to imitate what shall happen to the experimental gappy points in the PIV data in the next section. The images are then fed into the network to undergo training procedure.

Mutual impact of physical and data-based terms of the loss function on each other are now studied. The results of reconstruction are examined for similar conditions in six cases, using a data-based convolutional neural network and five physics-based convolutional neural networks. Figure 6 shows the variations of the network loss with the number of epochs for both physics-based and data-based terms. The two loss terms do not converge with similar patterns. The physical term dominates the loss function, due to amplification of errors in numerical calculation of derivatives. The data-based loss is far more well behaved at early epochs and converges when the physical term still shows large fluctuations. Figure A1 in the appendix presents samples of the reconstructed images.

Velocity field reconstruction is evaluated for six different configurations of $(\alpha_p, \alpha_d)$ to examine the contribution of physical and data-based terms of the loss function on the accuracy of the model. This includes a pure data-based CNN $(\alpha_p = 0, \alpha_d = 1)$, termed as the baseline configuration, and five physics-based CNNs with different weight configurations, namely $(\alpha_p = 0.01, \alpha_d = 1)$, $(\alpha_p = 0.1, \alpha_d = 1)$, $(\alpha_p = 1, \alpha_d = 1)$, $(\alpha_p = 1, \alpha_d = 10)$ and $(\alpha_p = 1, \alpha_d = $ 100). Figure 7 presents the scatter plot of prediction-target values for velocity components and mass conservation residual for these six configurations. Each point represents a snapshot and if any of these points is completely positioned on the line, it means on average perfect reconstruction of that snapshot. Additionally, Table 1 presents *MAE* calculated for these quantities, as well as the relative change of *MAE* ($\epsilon$) for each physics-based configuration with respect to the baseline (pure CNN) configuration, defined as follows:

$$\epsilon(\phi) = 100 \frac{MAE_{baseline}(\phi) - MAE_{physics-based}(\phi)}{MAE_{baseline}(\phi)} \quad (14)$$

For each target quantity $\phi$, positive value of $\epsilon$ indicates superior accuracy of physics-based network compared to the baseline network, and vise versa.

Figure 7 indicates the importance of selecting optimal weight coefficients for the network accuracy in terms of velocity or mass conservation metrics. The cluster of points for horizontal velocity reconstruction is mostly distinct from the 45-degree line, while the velocity reconstruction in the vertical direction and the mass conservation residual could in some cases be very close to the target values, depending on the selected set of coefficients. Positive values of $\epsilon$ in Table 1 indicate the superiority level of the physics-informed network as compared to the baseline model. It is shown in Table 1 that inclusion of physical term in the loss function in most of cases (expect case 5) enhances reconstruction of the horizontal velocity component, which is the dominant component of velocity field for the DNS data. Case 6 $(\alpha_p = 1, \alpha_d = 100)$ leads to the best accuracy, with respect to the baseline results, in terms of both velocity components. The relative enhancement is 96.6% and 42.43% for the horizontal and vertical components, respectively. However, this is not necessarily aligned with improvement of mass conservation residual. An opposite trend is observed in terms of mass conservation residual (by 102.19% relative reduction with respect to the baseline model). Overall, considering both velocity components and mass conservation residual metrics, case 4 $(\alpha_p = 1, \alpha_d = 1)$ results in the best performance. This case is a win-win compromise, such that 21.74% enhancement in $MC$ reconstruction accompanies by 60.96% and 20.39% improvement of reproduction for the velocity components.

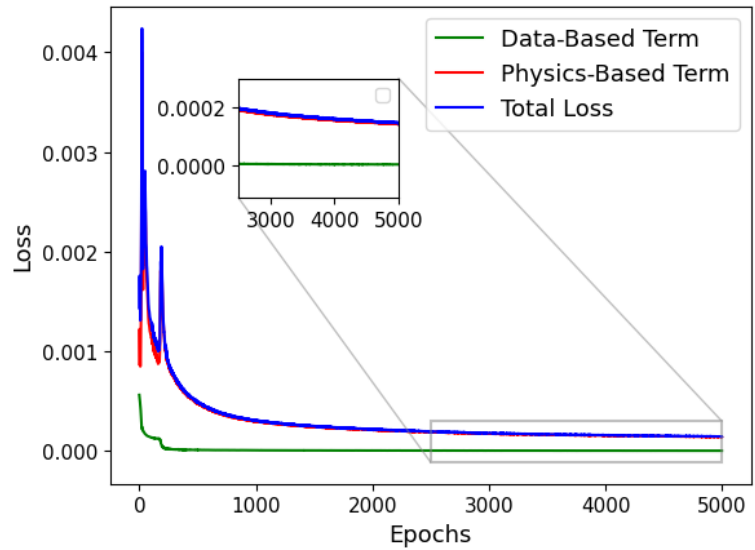

Figure 6 Convergence of the different loss terms during the training of PINN for the DNS data ($\alpha_P = 1, \alpha_d = 1$).

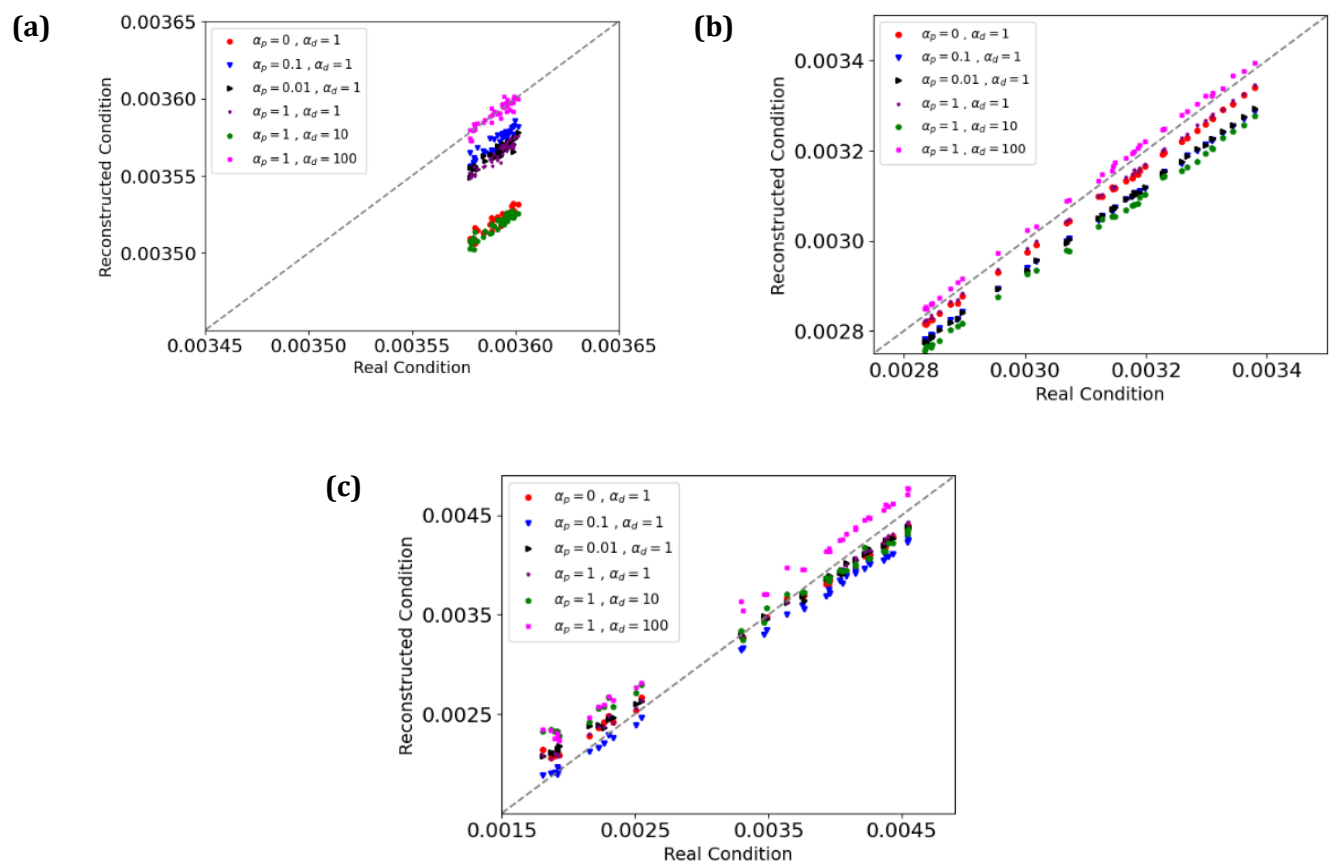

Figure 7 Comparison of model prediction and actual values for reconstruction of DNS data, in terms of: (a) horizontal velocity component; (b) vertical velocity component; (c) mass conservation residual.

Table 1 Accuracy of different physics-based configurations compared to the baseline network for the DNS data (the bold numbers in each column indicate the minimum error or the maximum percentage of improvement with respect to the baseline)

| Case | $\alpha_p$ | $\alpha_d$ | Horizontal velocity | | Vertical velocity | | Mass conservation residual | |
|---|---|---|---|---|---|---|---|---|
| | | | $MAE(u) \times 10^5$ | $\epsilon(\phi)$ % | $MAE(v) \times 10^5$ | $\epsilon(v)$ % | $MAE(MC) \times 10^5$ | $\epsilon(MC)$ % |
| 1 (baseline) | 0 | 1 | 7.07 | — | 3.04 | — | 13.20 | — |
| 2 | 0.1 | 1 | 1.89 | 73.26 | 7.20 | -136.84 | 16.75 | -26.89 |
| 3 | 0.01 | 1 | 2.58 | 63.50 | 7.26 | -138.81 | 12.67 | 4.01 |
| 4 | 1 | 1 | 2.76 | 60.96 | 2.42 | 20.39 | **10.33** | **21.74** |
| 5 | 1 | 10 | 7.30 | -3.25 | 9.05 | -197.69 | 20.31 | -53.86 |
| 6 | 1 | 100 | **0.24** | **96.60** | **1.75** | **42.43** | 26.69 | -102.19 |

Figure 7 and Table 1 also indicate that the quality of velocity field reconstruction does not solely depend on the ratio of $\alpha_p$ to $\alpha_d$, while it is also dependent to the value of these coefficients as well. For instance, if case 6 with $\alpha_p/\alpha_d = 0.01$, is compared with case 3 with the same weights ratio, the former shows a better accuracy in terms of mass conservation residual, but a lower accuracy for both velocity components.

## 4-2- PIV Data Reconstruction

The present section examines PINN model performance in reconstruction of the PIV data acquired at the rotor exit region of a centrifugal turbomachine. The images in this dataset all have $36 \times 36$ vector maps, while some points in each snapshot have missing velocities. Figure 8 illustrates histograms of four sample snapshots after outlier detection. Some of the velocity data (less than 1%) are outliers and should be removed in the pre-processing step. Among 2500 PIV snapshots, 90% of the normalized dataset is used for the training phase (2250 snapshots), and the rest for the test phase (225 snapshots). The subsequent pre-processing steps for the network's input are as presented in section 2-2, which are graphically illustrated in Figure 9 for one randomly selected snapshot. Fig. A2 in the appendix shows samples of reconstructed PIV data.

The network is typically converged after about 20000 epochs, which takes approximately 1.5 hours of training time on an NVIDIA Tesla P100 GPU available through Google Colab Pro. Once the model is trained, the deployment phase is highly efficient. The time required for a forward pass of the network computations on each new, unseen PIV snapshot is less than one second. This rapid processing time makes our approach quite applicable as a postprocessor in research studies.

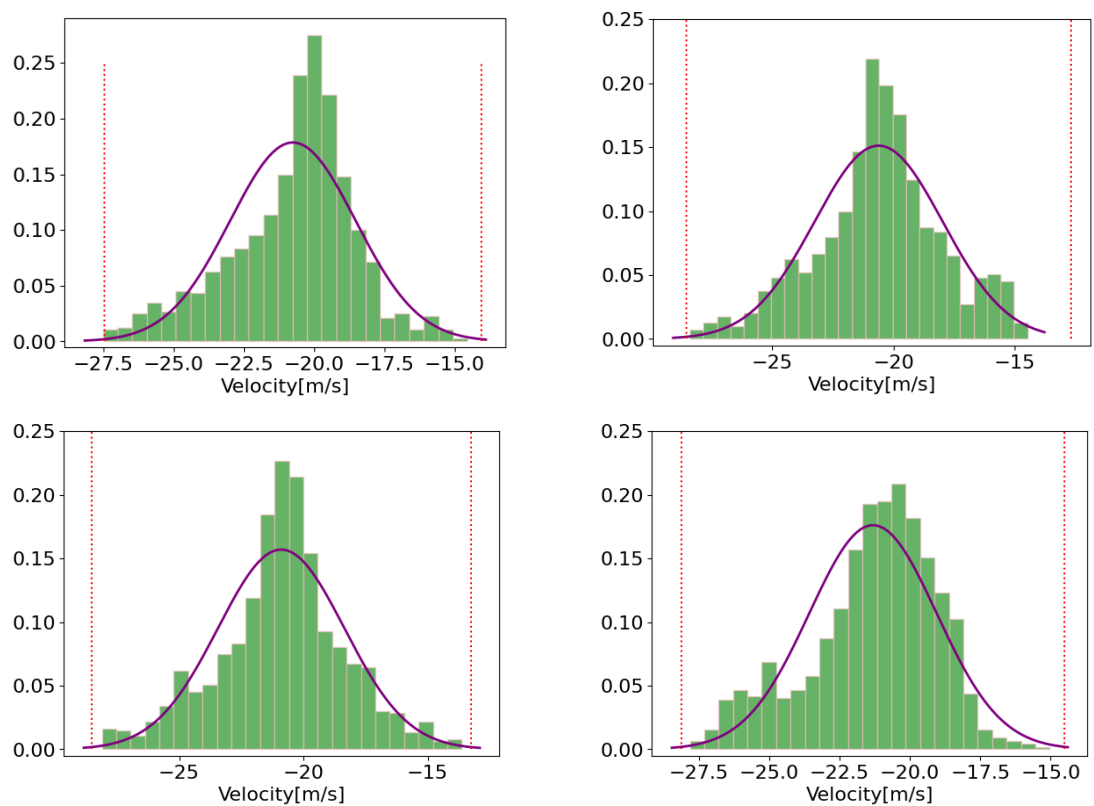

Figure 8 Histogram of instantaneous velocity for 4 random PIV snapshots after outlier removal. Red dash lines show the proposed band by z-score criteria.

It was shown in the previous section that a loss function with a mass conservation term, with suitable configuration of weight coefficients, could both improve DNS data reconstruction and make reconstructed data comply with the mass conservation equation. This was expected since the physical term was also used beforehand in the simulation step as a conservation law. To examine the same problem for the PIV data, Figure 10 illustrates the results of the physics-informed network for five different configurations of weight coefficients (same as section 4-1) and also compares them with the baseline pure CNN network. Table 2 presents the statistical evaluation of these results in terms of *MAE* and $\epsilon$. The point in Table 2 is the general enhancement of reconstruction accuracy for the horizontal velocity (as the dominant velocity component in this PIV data), in addition to improvement of mass conservation residual. The only exception is case 5 in which horizontal velocity component is almost unaffected by PINN model ($\epsilon_u = -0.52\%$). So, for such cases, in addition to better conformity of the model to the physical principles, the statistical accuracy of velocity field reconstruction is also improved.

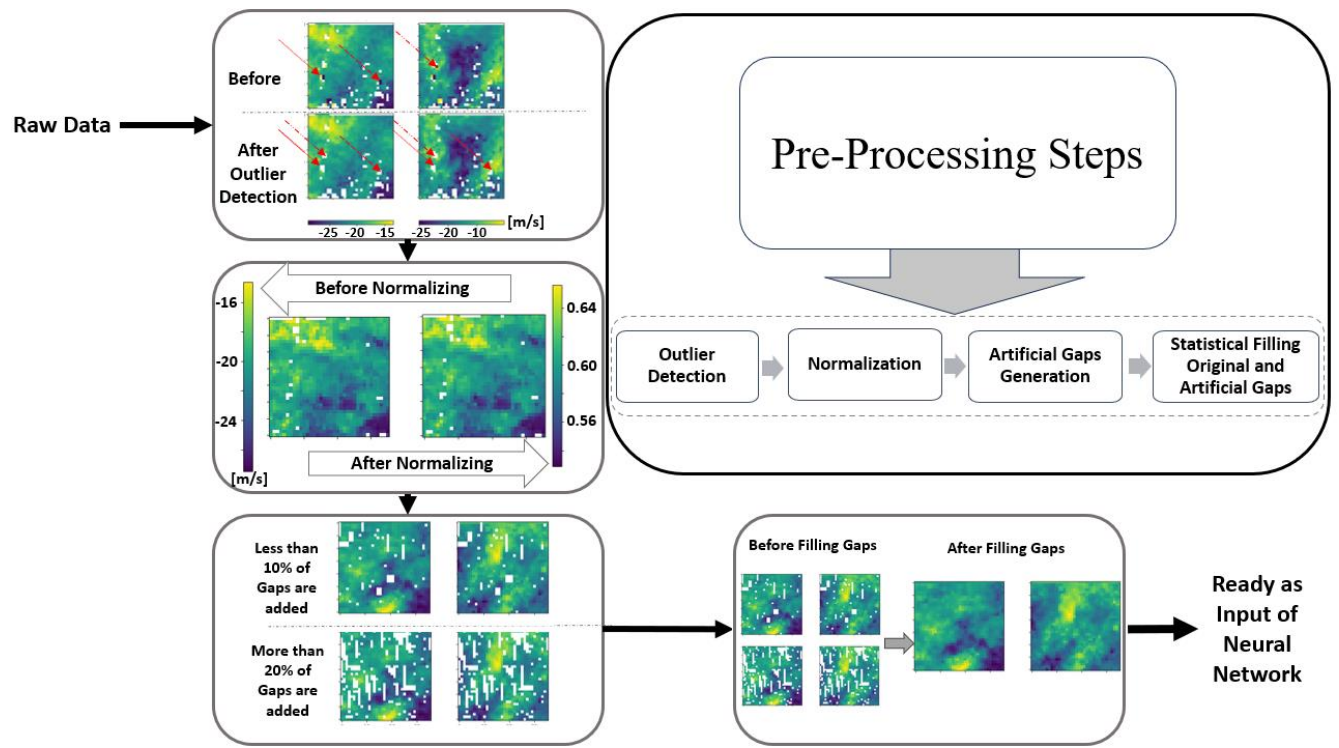

Figure 9 Representation of pre-processing steps for the PIV Data.

Unlike the DNS data, the PIV data is less sensitive to weighting coefficients, but derivatives are more difficult to reconstruct or optimize due to experimental noise and uncertainties. Table 2 shows that the worst case in terms of mass conservation metric is when the physical term is not present in the loss function (the baseline model). This does not mean that as weight coefficient corresponding to the physical term increases, reconstruction of mass conservation residual also enhances. Among the five configurations of weight coefficients, case 6 with the lowest weights ratio of $\alpha_p/\alpha_d = 1/100$ leads to the highest performance in terms of mass conservation and both velocity components. This ratio was calculated based on turbulence scale analysis too. A larger weight ratio highlights the physical term but due to amplification of experimental uncertainties in the derivative operation, such augmented contribution in the physical term does not necessarily enhance the reconstruction accuracy. This issue is also shown in Figure 10-c, that despite some displacement of cluster of points toward the 45-degree line, it is not possible to strictly satisfy the mass conservation law, due to such uncertainty amplifications. It should be noticed that during the training process, the network aims to optimize the overall loss, which includes both the data-driven and physics-based components, rather than each individual component separately. Consequently, there is a trade-off between implementation of physical loss and obtaining some enhancement in the reconstruction accuracy on the one side, and trying to give an improved reproduction of the third velocity gradient and more adherence to the continuity equation on the other side. For DNS data, the error margin is significantly lower, resulting in a more pronounced adherence to the continuity equation. This allows for more consistent tuning of the network based on both data-driven and physics-based loss terms. As previously shown in Figure 7, adjusting the weighting coefficients ($\alpha_d$ and $\alpha_p$) for the DNS data effectively balances accurate velocity reconstruction with adherence to the continuity equation.

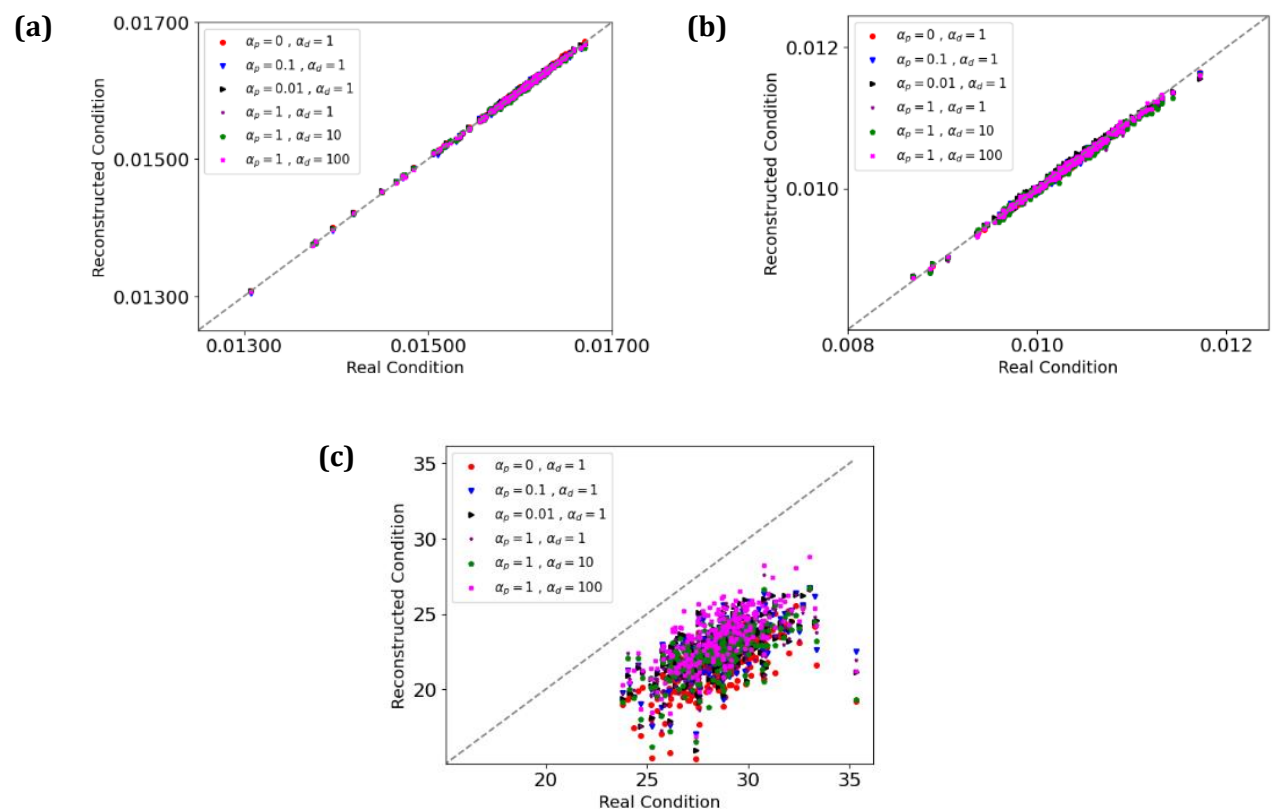

Figure 10 Comparison of model prediction and actual values for reconstruction of PIV data, in terms of: (a) horizontal velocity component; (b) vertical velocity component; (c) mass conservation residual.

Table 2 Accuracy of different physics-based configurations compared to the baseline network for the PIV data (the bold numbers in each column indicate the minimum error or the maximum percentage of improvement with respect to the baseline)

| Case | $\alpha_p$ | $\alpha_d$ | Horizontal velocity | | Vertical velocity | | Mass conservation residual | |
|---|---|---|---|---|---|---|---|---|
| | | | $MAE(u) \times 10^5$ | $\epsilon(\phi)$ % | $MAE(v) \times 10^5$ | $\epsilon(v)$ % | $MAE(MC)$ | $\epsilon(MC)$ % |
| 1 (baseline) | 0 | 1 | 1.92 | — | 2.55 | — | 7.05 | — |
| 2 | 0.1 | 1 | 1.59 | 17.18 | 3.20 | -25.49 | 5.86 | 16.87 |
| 3 | 0.01 | 1 | 1.64 | 14.58 | 2.83 | -10.98 | 5.89 | 16.45 |
| 4 | 1 | 1 | 1.62 | 15.62 | 3.59 | -40.78 | 6.00 | 14.89 |
| 5 | 1 | 10 | 1.93 | -0.52 | 3.96 | -55.29 | 6.09 | 13.61 |
| 6 | 1 | 100 | **1.37** | **28.64** | **2.52** | **1.17** | **5.09** | **27.80** |

Although based on the uncertainty level of any other data, a closer match of mass conservation residuals with the 45-degree line might be possible by tuning $\alpha_p$ and $\alpha_d$, this should be only attempted if any further participation of the physical term would not cross the uncertainty limit of the velocity components. This issue is studied for the current PIV data and the reconstruction error is compared with the experimental uncertainty. There are two types of uncertainties in the experimental data, namely random errors and bias errors. Random errors are due to non-systematic measurement characteristics, such as optical noise, and they can be reduced by repeating the experiment with stronger emphasis on the standard requirements and procedures. Bias errors are systematic and predictable. Errors due to particle lag effect, light refraction and displacement estimation are the most important bias errors in PIV measurement, which can be controlled by suitable selection and utilization of PIV components. The maximum bias error for the in-plane velocity components in the current data was estimated to be 1.5% [43].

Figure 11 presents the histogram of velocity components reconstruction based on three different configurations of weight coefficients, i.e. cases 3 and 6, and the baseline (pure data-based) model. It also compares the distribution with the normalized error band (red dashed line) deduced from experimental uncertainty. Quantitatively, Table 3 represents percentage of reconstructed vectors with error less than the experimental uncertainty, for each case in Figure 11. Table 3 indicates that the baseline model, reconstructs the PIV data such that about 90% of horizontal velocity components are within the experimental error band. By adding the physics-based term to the loss function, this percentage just reduces about 4-5%, while at the same time Table 2 reveals about 16% and 28% improvement in the reconstruction results of cases 3 and 6, in terms of the mass conservation residual, and about 15% and 29% enhancement in terms of horizontal velocity component. Therefore, limited excess from the uncertainty band in velocity reconstruction could enhance the reconstruction results, both in terms of data-based and physics-based metrics. However, this should be carried out by controlled adjustment of the contribution of physical term, because this derivative-based component can significantly amplify numerical instabilities resulting in deterioration of model convergence and accuracy.

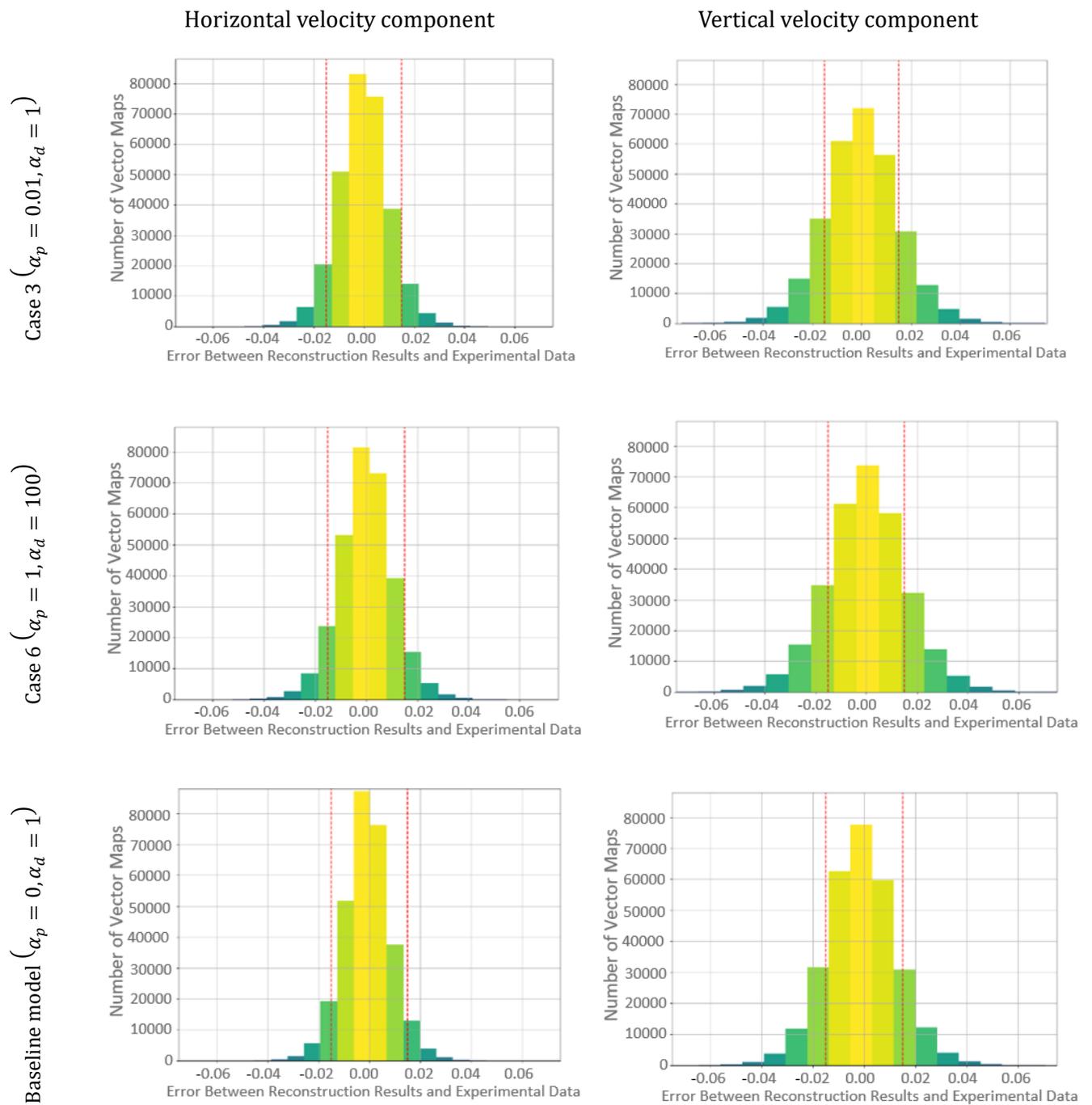

Figure 11 Histogram of reconstruction errors for horizontal (left column) and vertical (right column) components of velocity, based on three different configurations of weight coefficients

Table 3 Percentage of reconstructed vectors with error less than the experimental uncertainty.

|  | Case 3 $(\alpha_p = 0.01, \alpha_d = 1)$ | Case 6 $(\alpha_p = 1, \alpha_d = 100)$ | Baseline model $(\alpha_p = 0, \alpha_d = 1)$ |
|---|---|---|---|
| Horizontal velocity | 86.62% | 85.28% | 89.74% |
| Vertical velocity | 70.28% | 68.44% | 74.66% |

In comparison with the PINN approach, implementation of conventional statistical and algebraic methods for missing data imputation is more straightforward. In our previous study [15], we evaluated six conventional imputation methods for filling missing PIV data: gappy proper orthogonal decomposition (GPOD), nearest-neighbor interpolation, triangulation with linear interpolation (TLI), uniform-weighted moving average interpolation, Gaussian-weighted moving average interpolation, and moving median interpolation. Comparison of performance of these methods in reconstruction of PIV data with two machine learning models (support vector regression and neural network) indicated that the machine learning models performed superiorly or comparably to the TLI method and significantly outperformed the moving average/median methods. GPOD and nearest-neighbor interpolation were less successful due to the low frequency of PIV measurements or extensive clustered gappy regions in the flow domain. While these statistical methods are simpler and faster tools for velocity reconstruction, they do not ensure the physical validity of the reconstructed data. In contrast, the present PINN methodology, once trained, offers rapid deployment with a forward pass computation time of less than one second per new PIV snapshot. It provides the added value of physical validity alongside statistical accuracy.

To evaluate the generalizability of our network, we would ideally require experimental data for a flow field with a similar level of complexity to our current study. In other words, the consistency of the reported improvements across differenty types of fluid dyanamics datasets largely depends on the similarity of the challenges and flow regimes between the current study and other fluid dynamics problems, as well as the extent and size of invalidated data (gap regions). However, obtaining such experimental data was not feasible due to the lack of open access to large PIV datasets for complex fluid dynamics cases. Consequently, in the same way as almost all research in the AI community, we relied on CFD data for a simple benchmark flow to evaluate model performance, accuracy, and generalizability. Meanwhile, the aim of this paper was not to present a model that claims to be valid over a wide range of data sets. The prime target was to bring into the attention of the research community the intricate points in the trade-off between adhering to the original measured data or trying to compromise this reproduction within its error band, for satisfying the governing equations. This is a growing and state-of-the-art field of research, and while we do not claim that our approach is the definitive solution or it has no limitations, we believe it represents a significant step forward in addressing the contribution of physics-informed machine learning in experimental fluid dynamics scope.

# V) Conclusion

The performance of physics-informed convolutional neural networks for the reconstruction of experimental phase-resolved data around a centrifugal turbomachine was investigated. The objective was to impute PIV data in the gappy regions such that they comply with a trade-off between the mass conservation equation and the conventional velocity field statistics. This was carried out by proposing a convolutional auto-encoder network with a physics-based loss function relying on mass conservation residual. The proposed approach in the present study allows applying PINN reconstruction in a way that is suitable for studies relying on planar PIV measurements. We believe this methodology presents an accessible and efficient alternative for data reconstruction, adhering maximally to physical laws, particularly for research teams that may not have access to modern measurement techniques such as time-resolved tomographic PIV or double-plane SPIV.

DNS data of flow past a cylinder, as a physics-based data, was initially used to study the performance of the proposed PINN model. The model was then evaluated based on the planar PIV data, to further investigate its performance on a more challenging dataset involving experimental uncertainties. The results for both DNS and PIV datasets demonstrated that optimal incorporation of a physics-based equation into the loss function not only improves the overall agreement with the mass conservation equation, but also results in more accurate reconstruction of velocities.

The impact of weight coefficients for the physical and data-driven terms in the loss function was examined based on the accuracy of reconstructed data. Reconstructed data exhibited a noticeable movement towards satisfying mass conservation equation after adjusting weight coefficients in the DNS data. However, for the PIV data, although the addition of physical terms in the loss function proved to be beneficial, there was still a considerable gap to achieve complete reproduction of the physical equations. One possible reason is the existence of experimental uncertainty in the measured velocity field, which can be considerably amplified by implementation of derivative-based physical principles (like mass conservation law) to the loss function. Another reason could be the unavailable velocity gradients in the out-of-plane direction, which enforce a two-dimensional mass conservation residual to the loss function. Furthermore, the nature of PIV data in a turbomachine involves complex patterns, making network convergence significantly more challenging as compared with the DNS data of flow past a cylinder. Nevertheless, the results indicated that by controlled adjustment of weight coefficients, the network managed to guide the snapshots towards satisfying the mass conservation equation while maintaining a limited range of velocity reconstruction errors. Compared to data-driven model, the PINN network enhanced reconstruction quality by about 28% and 29% in terms of mass conservation residual and dominant velocity component, respectively, in expense of 5% increase in the number of vectors with reconstruction error larger than the uncertainty band.

## Acknowledgements

Authors extend their sincere gratitude to Mr. Ali Shadmani for his dedicated efforts, which significantly contributed to the valuable PIV data utilized in this article.

## conflict of interest statement

The authors have no conflicts to disclose.

## Author contributions

The contributions of authors in this study are as follows:

- Maryam Soltani: Methodology; Modeling; Formal analysis; Writing – original draft.
- Ghasem Akbari: Conceptualization; Methodology; Review & editing; Supervision.
- Nader Montazerin: Conceptualization; Methodology; Review & editing; Supervision.

## Data availability statement

The data that support the findings of this study are available from the corresponding author upon reasonable request. The source codes and sample data will be included after acceptance of the article through GitHub repository.

## Appendix: Samples of reconstructed velocity fields

Figures A1 and A2 illustrate samples of reconstructed velocity fields, based on DNS and PIV data, respectively. Figure A1 has four columns, in which the first two left columns represent reconstruction of horizontal and vertical velocity components for one random DNS snapshot, based on various configurations of loss weights. The other two columns illustrate the error between target and reconstructed images. Figure A2 indicates the results of PIV data reconstruction for three random snapshots in horizontal and vertical directions, based on different weight coefficients. The last row, in both Figures A1 and A2, belongs to the baseline (data-based) network.

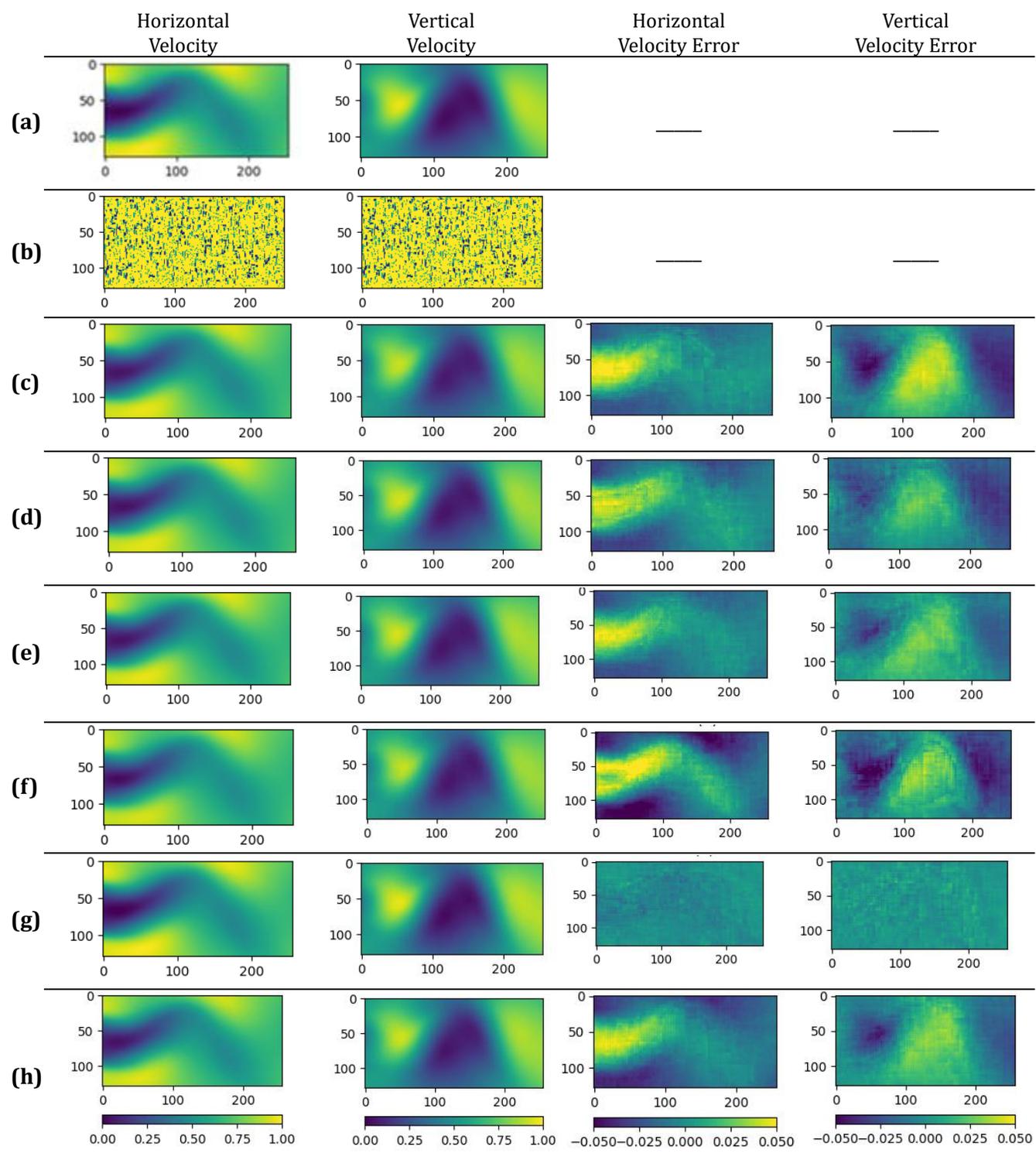

Figure A1. Samples of Reconstructed velocity field for DNS Data: a) target field; b) artificial gaps; c) $\alpha_p = 0.1, \alpha_d = 1$; d) $\alpha_p = 0.01, \alpha_d = 1$; e) $\alpha_p = 1, \alpha_d = 1$; f) $\alpha_p = 1, \alpha_d = 10$; g) $\alpha_p = 1, \alpha_d = 100$; h) $\alpha_p = 0, \alpha_d = 1$.

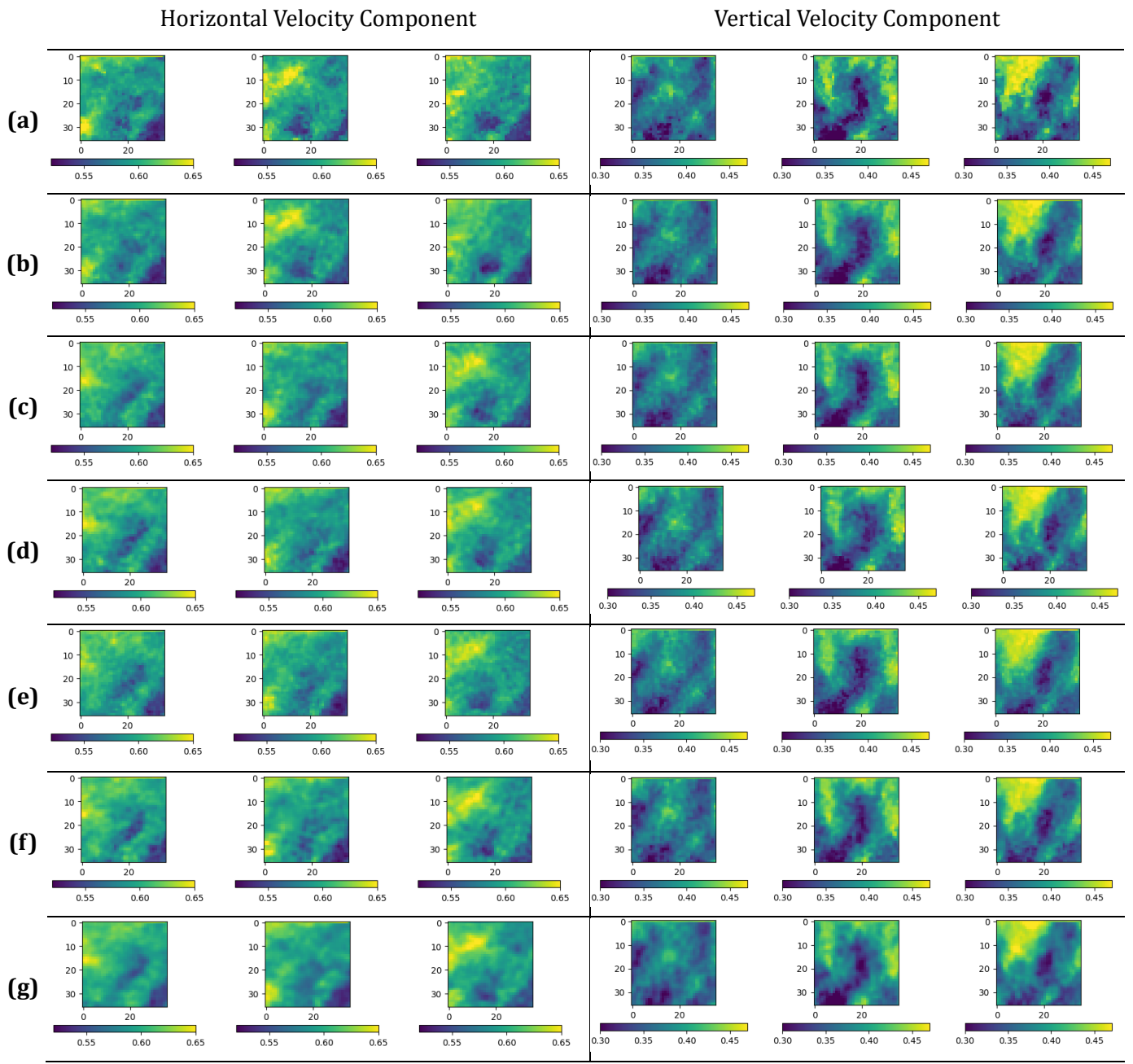

Figure A2. Three samples of reconstructed PIV Snapshots: a) Target field; b) $\alpha_p = 0.1, \alpha_d = 1$; c) $\alpha_p = 0.01, \alpha_d = 1$; d) $\alpha_p = 1, \alpha_d = 1$; e) $\alpha_p = 1, \alpha_d = 10$; f) $\alpha_p = 1, \alpha_d = 100$; g) $\alpha_p = 0, \alpha_d = 1$.